\documentclass[11pt,twoside]{article}

\usepackage{asp2010}

\resetcounters

\begin{document}

\title{Theoretical approaches to the physics of spectral line polarization}

\author{Luca Belluzzi
\affil{Instituto de Astrof\'isica de Canarias, E-38205 La Laguna, Tenerife, Spain}}

\begin{abstract}
Due to the continuous developments in polarimetric instrumentation, which will 
become even more dramatic in the near future with the availability of new 
generation solar telescopes, we are now severely confronted with a variety
of new detailed observations of high diagnostic potential, whose interpretation
requires a firmly established theoretical framework.
In this contribution, I review the fundamental physical processes that 
underlie the generation and transfer of polarized radiation in stellar 
atmospheres, and I discuss the present status of the theoretical schemes 
now available, pointing out their main successes and limitations.
I also present some ideas about the theoretical improvements that I
consider necessary to achieve a correct interpretation of the complex
phenomenology shown by polarimetric observations, focusing particularly on
the second solar spectrum, which can be considered as one of the most important
test benches of the theory.
\end{abstract}

\section{Introduction}

Polarized radiation is expected whenever the interaction process between matter 
and radiation is characterized by any kind of symmetry-breaking \citep[see, 
for example, the review by][]{Cas07}. 
The symmetry-breaking can be intrinsic to the interaction itself, or due to 
external causes (e.g., presence of a magnetic field).
In order to get more insights on this fundamental statement, let us consider a 
quantum transition between an upper level $J_u$, and a lower level $J_{\ell}$, 
$J$ being the quantum number associated with the total angular momentum operator.
According to quantum mechanics, the atomic system in the upper (or lower) level
can occupy any of the $(2J+1)$ possible $M$ magnetic substates, which are 
degenerate if no magnetic field is present.
The radiation emitted by the various transitions between the $M_u$ substates 
of the upper level and the $M_{\ell}$ substates of the lower level (generally 
referred to as Zeeman components), is characterized by well defined 
polarization properties which depend on the angle between the emission 
direction and the quantization axis, and on the value of 
$\Delta M \equiv M_u - M_{\ell}$. 
In the electric dipole approximation the only allowed transitions are those 
with $\Delta M=0$ ($\pi$ components), and those with $\Delta M = \pm 1$ 
($\sigma_{\rm b}$ and $\sigma_{\rm r}$ components, respectively).
The polarization properties of the $\pi$ and $\sigma$ components emitted along 
various directions are schematically shown in Fig.~1.
\begin{figure}[!t]
\centering
\includegraphics[width=0.5\textwidth]{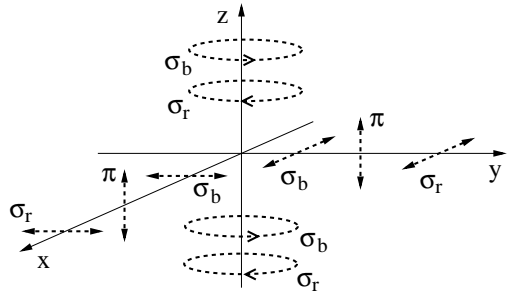}
\caption{Polarization properties of the radiation emitted by the different 
Zeeman components. The $z$ axis is the quantization axis.}
\end{figure}

In a de-excitation process, each Zeeman component contributes to the emitted
radiation with a particular weight which takes into account the relative
population of the upper $M$-sublevel of the component, the relative strength
of the component, and the emission direction.
If the atomic system is naturally (isotropically) excited, then the
upper magnetic sublevels are equally populated, and it can be shown that
the weights of the various Zeeman components assume values such that, if no 
magnetic field is present, the total radiation emitted in any direction is 
completely unpolarized.

There are two basic scenarios (not mutually exclusive) where the different
polarization properties of the Zeeman components manifest themselves in
emission of polarized radiation.
The first scenario corresponds to the possibility that the $M$ substates
may be separated in energy, so that the contributions of the various
components fall at slightly different wavelengths. In this case, even if the
the atomic system is naturally excited, the emitted radiation has polarization
properties varying with wavelength. 
The most common circumstance for this condition to occur is when the atomic 
system interacts with an external magnetic field ({\it Zeeman effect}).
The second scenario corresponds to the possibility that the upper $M$ substates 
are unevenly populated, so that the various components no longer contribute to 
the emitted radiation with those particular weights that make the total
polarization vanish.
This condition generally occurs whenever the atomic system is excited 
through a physical process which, for any reason, is not spatially isotropic.
An atom whose magnetic sublevels are not evenly populated and/or are 
characterized by well-defined phase relations is said to be polarized.
The radiation emitted by a polarized atomic system is in general polarized, as 
it can be shown through the following clarifying example.
Let us consider a two level atom, the lower level having $J_{\ell}\!=\!0$ and 
the upper level having  $J_u\!=\!1$, and let us suppose to excite it with a 
collimated unpolarized radiation beam propagating along the quantization axis. 
In the presence of such an incident radiation field, it can be shown that 
only the $M_u\!=\!\pm 1$ magnetic sublevels will be populated, so that only the 
$\sigma$ components will contribute to the emitted radiation. 
From Fig.~1 it is then clear that if the radiation emitted perpendicularly to 
the incident beam is considered, this will be totally linearly polarized 
perpendicularly to the scattering plane.
It should be observed that this is basically the physical mechanism responsible 
for the linearly polarized spectrum of the solar radiation coming from 
quiet regions close to the limb (the so-called {\it Second Solar Spectrum}).
Of course, in the solar atmosphere the radiation field impinging on the atoms
is not perfectly collimated along the local vertical, nevertheless it has a 
certain degree of anisotropy that provides for the radiation scattered at 
90$^{\circ}$ a linear polarization degree of the order of 1\%.
Polarization produced through resonance scattering processes like the one 
here described is usually referred to as {\it scattering polarization}.

The quantum theory of the Zeeman effect was developed in the first half of 
the last century, and we can safely say that today it is absolutely well 
established.
Indeed, the difficulties related to the interpretation of spectropolarimetric 
signals produced through this mechanism are essentially due to the complexity 
of the line formation problem in the solar atmosphere, the physical 
origin of the polarization being, on the other hand, absolutely clear.
On the contrary, as soon as the first scattering polarization signals, 
detected at the solar limb since the early 1940s \citep[see][]{Red41}, could 
be observed with a sufficient accuracy, it was immediately clear that the 
theoretical interpretation of the peculiar profiles shown by several of them 
would have required a substantial improvement of our understanding of the 
physics of resonance scattering.
The incomplete comprehension of this fundamental physical process has become 
particularly evident during the last decades when, thanks to the development of 
polarimeters like ZIMPOL \citep[see][]{Pov95}, able to reach sensitivities 
of the order of 1 part over 10$^5$, many new apparently ``enigmatic'' 
scattering polarization signals have been detected \citep[see][]{Ste97b}. 
Given the high number of physical ``ingredients'' involved in the generation 
of scattering polarization signals, and given their consequent enormous 
diagnostic content, since the 1970s their interpretation has been representing 
one of the most important and exciting theoretical challenges in the field of 
solar polarimetry.

Though our understanding of the physics of resonance scattering has 
substantially improved during the last 30 years, and a robust theoretical 
scheme, based on the principles of quantum electrodynamics has been developed
\citep[see][hereafter LL04]{Lan04}, and successfully applied for the 
interpretation of a wide range of signals 
\citep[see][for a recent review]{JTB09}, our comprehension of the 
rich phenomenology shown by scattering polarization signals remains 
rather fragmentary.

In the present contribution I will review some of the most important 
theoretical approaches to the physics of scattering polarization developed so 
far. The aim of the paper is to point out strengths and limitations of the 
various approaches by reviewing the physical processes and phenomena they are 
able to describe, and by discussing the most important hypotheses and 
approximations they are based on (for a detailed derivation of each approach 
the reader will be referred to the original papers or to more 
exhaustive monographs).
Examples of the main successes of each theory, as well as a discussion of 
unsolved problems, will also be presented.
Of course, the review is by no means exhaustive, and I apologize in advance 
to all whose work is not mentioned here.

\section{The second solar spectrum}
The linearly polarized solar limb spectrum, today known as the second solar 
spectrum, is probably the clearest manifestation of scattering polarization in 
astrophysics, and it is no coincidence that it played, and is still playing a 
fundamental role in the development of the theory of polarization.
Its importance can be clearly highlighted through a few remarkable examples 
of signals whose interpretation required fundamental advances in our 
understanding of the physics of resonance scattering.

The interpretation of the peculiar polarization pattern observed across the H 
and K lines of Ca~{\sc ii} (and, similarly, across the D$_1$ and D$_2$ lines 
of Na~{\sc i}), which shows a sign reversal between the two lines, required 
the comprehension of the fundamental role played by quantum interferences 
(or coherences) between different quantum levels, as first showed by 
\citet{Ste80}.
In 1997, Stenflo explained the triplet-peak structure of the second solar 
spectrum of the Ba~{\sc ii} D$_2$ line in terms of hyperfine structure (HFS), 
pointing out how this physical aspect, generally negligible in the usual 
intensity spectrum, can be extremely important when polarization phenomena 
are taken into account \citep[see][]{Ste97a}.
In the same year, \citet{JTB97} pointed out the importance of lower level 
polarization, previously systematically neglected due to the qualitative 
idea that atomic polarization in long-lived levels should be always completely 
destroyed by depolarizing collisions and magnetic fields.
By taking into account this new ingredient it has been possible to 
interpret the polarization signals observed in the Ca~{\sc ii} infrared 
triplet line at 8662~{\AA} \citep[see][]{Man03}, as well as in the 
Mg~{\sc i} b-lines \citep[see][]{JTB99,JTB01}.

There are now clear evidences that the interpretation of a large class of 
signals (like the one observed in the Ca~{\sc i} line at 4227~{\AA}) requires 
the effects of partial redistribution in frequency to be taken into account. 
The development of theoretical frameworks able to account for these effects is 
nowadays one of the hottest topics in this research field.
Laboratory experiments are now being performed in order to clarify various 
aspects of this spectrum, like the presence of line-integrated linear 
polarization in the D$_1$ line of sodium and potassium 
\citep[see][]{Tha06,Tha09}, and in order to verify the main results derived 
from the theoretical approaches developed so far.

The importance of the second solar spectrum is not limited to single signals,
but also concerns some of its general properties, as recently pointed out by 
\citet{Bel09a,Bel10}.
Using the atlas ``The Second Solar Spectrum'' \citep{Gan00,Gan02,Gan05}, 
\citet{Bel09a} identified the strongest signals of this spectrum, and divided 
them into five different classes according to the shape of their profile. 
Through such a systematic analysis, the authors discovered a series of 
spectroscopic properties common to the spectral lines responsible for such 
signals, properties that have been summarized into the following three 
empirical laws:
\begin{enumerate}
	\item{{\bf First law:} {\it the transitions producing the strongest 
		polarization signals of the second solar spectrum are either 
		resonance transitions, or subordinate transitions whose lower 
		level is the upper level of a resonance transition producing a 
		strong polarization signal.}}
	\item{{\bf Second law:} {\it all the strong polarization signals that 
		are produced by spectral lines having a small equivalent width 
		(i.e. lines having $W_{\lambda}/\lambda < 20$~F) are of type 
		``S''\footnote{All the signals whose profile shows a single 
		sharp peak have been classified as ``S'' signals.}.}}
	\item{{\bf Third law:} {\it the spectral lines producing the strongest
		 polarization signals ($Q/I > 0.17\%$) are due to transitions 
		 having either $\Delta J \equiv J_u - J_{\ell}=+1$ or 
		 $\Delta J=0$.}}
\end{enumerate}
The theoretical interpretation of these laws is already under investigation, 
and it will probably represent a further important test bench for the 
theoretical approaches that will be developed in the future.

\section{A unifying theoretical approach in the formalism of quantum electrodynamics}

The unifying, self-consistent theoretical approach to the physics of 
polarization, formulated by \citet{Lan83} within the framework of the 
density-matrix formalism, starting from the principles of quantum 
electrodynamics, represents one of the most important theoretical 
accomplishments obtained so far in this research field.
A clear and detailed derivation of this theoretical scheme can be found in
LL04, and will not be repeated here. 
In this section, in order to point out strengths and weaknesses of this 
approach, I will present only the fundamental equations the theory is based on, 
discussing their physical meaning, and focusing the attention on the most 
important hypotheses and approximations that are introduced 
\citep[see also][]{Cas07,JTB90}.
I will then review some of the most important successes of this theory, and I
will discuss its main limitations and their consequences for the interpretation 
of particular classes of signals.

Other theoretical approaches, developed with different assumptions and 
suitable for investigation of those signals which cannot be interpreted 
under the approximations of the scheme described here, will be 
presented in the next section.

\subsection{The density operator}

The first problem to be addressed in developing a quantum theory for scattering
polarization is to find a suitable way to describe atomic polarization 
(i.e. the presence of population unbalances and/or of quantum interferences 
between pairs of magnetic sublevels).
A very good solution, as proposed by \citet{Bom78} and by \citet{Bom80} 
is through the density operator, an extremely powerful tool for describing any 
physical system that is in a statistical mixture of states.
The density operator is defined by
\begin{equation}
	\hat{\rho} = \sum_{\alpha} p_{\alpha} \, |\, \psi_{\alpha} \rangle \, 
	\langle \psi_{\alpha}| 
	\; ,
\end{equation}
where $p_{\alpha}$ is the probability for the system to be in the (pure) 
dynamical state described by the vector $|\psi_{\alpha}\rangle$, and where the 
sum is extended to all the pure states in which the system can be found.
The action of the density operator is completely specified once its matrix 
elements, evaluated on a given basis of the Hilbert space associated to the 
quantum system, are known. Such density-matrix elements contain all the 
accessible information about the system; through them it is possible to 
describe the dynamical state of the system and its temporal evolution in a 
very compact way.
For an atomic system, the most natural basis for defining the matrix elements
of the density operator is the basis of the eigenvectors of the total angular 
momentum. On this basis the elements of the density matrix are given by
\begin{equation}
	\langle \alpha J M \, | \, \hat{\rho} \, | \, \alpha^{\prime} J^{\prime} 
	M^{\prime} \rangle = \, \rho \, (\alpha J M, \alpha^{\prime} J^{\prime} 
	M^{\prime}) \; ,
\end{equation}
where $J$ and $M$ are the angular momentum quantum numbers previously 
introduced, while $\alpha$ represents a set of inner quantum numbers.
As it can be shown (see, e.g., LL04), the diagonal elements represent 
the populations of the magnetic sublevels, while the off-diagonal elements 
describe the quantum interferences between different magnetic sublevels.
The expectation value of a given dynamical variable $O$ is given by (see, e.g., 
LL04)
\begin{equation}
\label{eq:trace}
	\langle O \rangle = \, {\rm Tr} \, ( \, \rho \, \hat{O} \, ) \; ,
\end{equation}
with $\hat{O}$ the quantum operator associated to the dynamical variable $O$.

\subsection{The statistical equilibrium and the radiative transfer equations}

The time evolution of the density operator associated to a physical system 
described by the Hamiltonian $H$ is given by the Liouville equation
\begin{equation}
	\frac{\rm d}{{\rm d}t} \, \rho(t) = - \frac{2 \pi {\rm i}}{h} 
	\big[ H, \rho(t) \big]  \; .
\end{equation}
If the physical system is composed by one atom and by the radiation field, then 
the Hamiltonian $H$ is the given by the sum $H = H_A + H_R + V$, with $H_A$ the 
atomic Hamiltonian (which eventually includes the interaction with an external 
magnetic field), $H_R$ the Hamiltonian of the radiation field (generally 
described within the formalism of the second quantization), and $V$ the 
interaction Hamiltonian 
\citep[for its explicit expression see, for example,][]{Coh77}.

When the total Hamiltonian can be expressed as the sum of an unperturbed 
Hamiltonian and an interaction Hamiltonian, it is generally useful to work in 
the so-called interaction picture.
In this representation, in fact, only the interaction Hamiltonian appears in 
the commutator in the right-hand side of the Liouville equation, which assumes the form
\begin{equation}
\label{eq:Liu}
	\frac{\rm d}{{\rm d}t} \rho_{\rm I}(t) = - \frac{2 \pi {\rm i}}{h} 
	\big[ V_{\rm I}, \rho_{\rm I}(t) \big]  \; ,
\end{equation}
where the label $I$ means that the corresponding quantity is defined in the 
interaction picture.
Equation~(\ref{eq:Liu}) has the formal solution
\begin{equation}
\label{eq:formal}
	\rho_{\rm I}(t)=\rho_{\rm I}(0)-\frac{2 \pi{\rm i}}{h}\int_{0}^{ \,t}
	\big[ V_{\rm I}(t^{\prime}),\rho_{\rm I}(t^{\prime}) \big] \, 
	{\rm d}t^{\prime}  \; .
\end{equation}
This equation can be used again to express the quantity 
$\rho_{\rm I}(t^{\prime})$ that appears in the integral. 
By repeating this procedure, a perturbative development of the formal solution 
of Eq.~(\ref{eq:Liu}) is obtained.
If the development is stopped at the $k$-th term, by substituting the quantity 
$\rho_{\rm I}(t^{\prime})$ in the integral with $\rho_{\rm I}(0)$, the 
matter-radiation interaction is said to be described at the $k$-th perturbative 
order.

The equation describing the time evolution of the expectation value of a given 
dynamical variable $O$ can be obtained calculating the time derivative of 
Eq.~(\ref{eq:trace}), and using Eq.~(\ref{eq:formal}).
After some algebra one obtains
{\setlength\arraycolsep{2pt}
\begin{eqnarray}
\label{eq:obs-evol}
\frac{\mathrm{d}}{\mathrm{d}t} \, O(t) & = & \mathrm{Tr} \bigg\{ \bigg( \frac{
        \mathrm{d}}{\mathrm{d}t} \, \hat{O}_{\mathrm{I}}(t) \bigg)
        \, \rho_{\mathrm{I}}(t)
        \bigg\} - \frac{2 \pi \mathrm{i}}{h} \mathrm{Tr} \bigg\{ \big[ 
	\hat{O}_{\mathrm{I}}(t),V_{\mathrm{I}}(t) \big] \,
        \rho_{\mathrm{I}}(0) \bigg\} \nonumber \\
        & & - \frac{4 \pi^{2}}{h^{2}} \mathrm{Tr} \bigg\{ \int_{0}^{\, t} 
	\Big[ \big[\hat{O}_{\mathrm{I}}(t),V_{\mathrm{I}}(t) \big], 
	V_{\mathrm{I}}(t^{\prime}) \Big] \, 
	\rho_{\mathrm{I}}(t^{\prime}) \, \mathrm{d}t^{\prime} \bigg\}  \; .
\end{eqnarray}}
This is an exact equation, and represents the starting point to derive the 
equations describing the evolution of the atomic system and of the radiation 
field resulting from their mutual interaction.
In particular, if in place of $O(t)$ and $\hat{O}_{\rm I}(t)$ one puts the 
density matrix elements associated to the atomic system, and the corresponding 
quantum operator, then the statistical equilibrium equations for the density 
matrix are obtained. 
If in place of $O(t)$ and $\hat{O}_{\rm I}(t)$ one puts the Stokes parameters 
describing the radiation field, and the corresponding quantum operators, then 
the radiative transfer equations are obtained.

The approximation that is now introduced is to substitute the quantity 
$\rho_{\rm I}(t^{\prime})$ that appears in the integral in the right-hand side of
Eq.~(\ref{eq:obs-evol}) with $\rho_{\rm I}(0)$, so that it can be extracted 
from the integral.
Since this is equivalent to performing the same substitution in the integral 
on the right-hand side of Eq.~(\ref{eq:formal}), this approximation is equivalent to 
consider a {\it 1}-st order perturbative development of the matter-radiation 
interaction, which means that only {\it 1}-st order processes (processes 
involving only one photon) can be described.

Within this approximation, scattering, which is intrinsically a {\it 2}-nd 
order process, has to be treated as a succession of independent {\it 1}-st 
order absorption and emission processes (i.e.\ in the limit of complete 
redistribution in frequency, CRD).
It should be observed that this is correct either in the presence of 
collisions that perturb the atom during the interaction with the radiation 
field to the point of completely relaxing the coherence of the scattering 
process, or when the radiation field illuminating the atom is spectrally flat 
(i.e.\ independent of frequency).
Since polarization phenomena are generally negligible when the first condition 
occurs (because of the effect of depolarizing collisions), the latter 
(usually referred to as the {\it flat-spectrum approximation}) is generally 
assumed when the discussed theoretical approach is applied. 
More precisely, this approximation requires the radiation field to 
be flat across spectral intervals larger than the Bohr frequency connecting 
pairs of magnetic sublevels between which quantum coherences are considered, 
and larger than the inverse lifetime of the various magnetic sublevels. 

The statistical equilibrium equations for the density matrix were first 
derived (under the previously discussed approximation) by \citet{Bom78} and 
\citet{Bom80}.
The equations describe the transfer and relaxation of populations and 
coherences due to absorption and emission processes, and due to the 
presence of a magnetic field (Hanle effect).
The same equations were then rederived, under the same approximation, in the 
seminal paper by \citet{Lan83}. 
In this work Landi Degl'Innocenti derived within the same formalism both the 
statistical equilibrium equations for the density matrix, and the radiative 
transfer equations for polarized radiation (described in terms of the four 
Stokes parameters), thus ``closing'' the Non-LTE problem in the presence of 
polarization phenomena (Non-LTE problem of the 2$^{\rm nd}$ kind).
The equations include the effects of a magnetic field (Zeeman and Hanle
effects), and of collisions (elastic and inelastic). 
A detailed derivation of these equations, as well as a complete discussion of 
the important aspect (here neglected) of velocity/density-matrix correlations, 
can be found in LL04.

\subsection{Strengths and successes of the theory}

The main strength of the theoretical scheme described above is clearly its
self-consis\-tency, since all the equations are derived within the same elegant 
formalism from the principles of quantum electrodynamics. 
Going to more specific aspects, particularly important strength points of the 
theory are the following:
\begin{itemize}
	\item{possibility to take into account lower level polarization;}
	\item{possibility to take into account coherences within the same 
		$J$-level (multi-level atoms) and between different $J$-levels
		(multi-term atoms);}
	\item{possibility to treat atoms with hyperfine structure;}
	\item{possibility to describe the interaction with a magnetic 
		field in every regime, from the Zeeman effect regime to the 
		complete Paschen-Back effect regime;}
	\item{possibility to take into account both the Zeeman and the Hanle 
		effects.}
\end{itemize}
The theory has been extensively and very successfully used for the 
determination of magnetic fields in filaments and prominences, 
where the flat-spectrum approximation is generally well justified
\citep[see, e.g.,][]{Bom81,Ath83,Bom94,JTB02}.
Thanks to the possibility of taking into account quantum coherences between 
different $J$-levels, through this theoretical scheme it is possible to 
reproduce the peculiar polarization patterns (see Sect.~2) observed 
in the second solar spectrum across the H and K lines of Ca~{\sc ii}, and 
across the D$_1$ and D$_2$ lines of Na~{\sc i} (see LL04).
It was within the framework of this theoretical approach that 
\citet{JTB99,JTB01} and \citet{Man03}, as mentioned in Sect.~2, first 
explained, in terms of lower level polarization, the ``enigmatic'' signals 
produced in the same spectrum by the Mg~{\sc i} b-lines and by the 
Ca~{\sc ii} infrared triplet, respectively.
The possibility of treating multi-level atoms has made this approach 
particularly attractive for investigating complex atoms.
Very interesting results have been obtained, for example, for the 
interpretation of the second solar spectrum of Ti~{\sc i} \citep{Man02} and 
Ce~{\sc ii} \citep{Man06}.

An interesting result, which could be obtained thanks to the possibility 
for the theory to model atoms with HFS, and to take into account both the Hanle 
and Zeeman effects in every magnetic regime, is the discovery of a differential 
magnetic sensitivity of the triple-peak signal shown in the second solar 
spectrum by the Ba~{\sc ii} D$_2$ line \citep[see][]{Bel06,Bel07}. 
Such differential magnetic sensitivity, which has great diagnostic 
potentialities, has already been confirmed by various observational campaigns 
\citep[see, e.g.,][]{Ram09,Lop09}.
It should be observed that the possibility of considering coherences between 
different $J$-levels, or between different HFS $F$-levels, in the presence of a 
magnetic field in the incomplete Paschen-Back effect regime allows the theory 
to account for quantum mechanisms like the level-crossing and the 
anti-level-crossing\footnote{This expression has been proposed by 
\citet{Bom80}.} effects \citep[see, e.g.,][]{Bel07}.

Within the framework of this theory, \citet{Bel09b} investigated the 
physical origin of the weak polarization signal observed by \citet{Ste00} 
in the second solar spectrum of the faint lithium doublet at 6708~{\AA}, and 
concluded, in disagreement with this observation showing a triplet-peak 
structure, that only a two-peak profile can be theoretically expected, and that 
only a one-peak profile should be measurable unless the temperature of the 
atmospheric layer the radiation comes from is sufficiently low.
On the other hand, improved observations (like those presented in Stenflo 2011,
and others which are still unpublished) seem to agree with the one-peak
profile predicted by \citet{Bel09b}. 
The theoretical result of \citet{Bel09b} may thus be considered as an example 
of a theoretical prediction that anticipated an observational finding, an 
important indication of the adequacy of the theory.
The soundness of the theory can be also appreciated in that it has allowed 
to predict new physical mechanisms like the so-called ``alignment-to-orientation
conversion mechanism'' (see LL04), or the ``atomic-polarization-induced 
Faraday pulsation effect'' \citep[see][]{Lan11}.

It should be finally pointed out that the results obtained by applying this 
theory to different multiplets are in good qualitative agreement with the 
third law formulated by \citet{Bel09a}. Other examples of successful 
applications of this theoretical scheme can be found in \citet{JTB09}.

\subsection{Limitations and open problems}

The main limitation of this theoretical approach is the impossibility to take 
into account effects of partial redistribution (PRD) in frequency.
While in the intensity spectrum such effects are important (mainly in the 
wings) only for a limited number of very strong resonance lines (e.g., 
Ca~{\sc ii} H\&K, Mg~{\sc ii} h\&k, H$\alpha$, Ly$\alpha$), in the second solar 
spectrum, as pointed out by several works, starting with the pioneering one 
by \citet{Omo72}, they seem to be the key ingredient to interpret most of 
the linear polarization profiles that \citet{Bel09a} have classified as ``M'' 
signals\footnote{According to this classification the ``M'' signals are all 
those signals which show positive polarizing peaks (or lobes) in the wings of 
the line, a depolarization approaching to the line center, and (eventually) a 
sharp polarizing peak in the line core.} (typical examples are the signals 
produced by the Na~{\sc i} D$_2$ line or by the Ca~{\sc i} line at 4227~\AA).

The open problems, however, are not limited to the interpretation of the 
``M'' signals in terms of PRD effects.
An interesting example of an unsolved problem concerning the second solar 
spectrum is the physical origin of the triplet-peak profile shown by 
the Sc~{\sc ii} line at 4247~{\AA}.
This signal is very similar to the one produced by the Ba~{\sc ii} D$_2$ line,
which has been clearly interpreted in terms of the HFS shown by two of the 
seven stable isotopes of barium \citep[see][]{Ste97a, Bel07}.
Given that the only stable isotope of scandium shows HFS, and noticing the 
similarities between the two lines in the intensity spectrum, it was believed 
that this signal could also be interpreted in terms of HFS\footnote{It should 
be observed that only 18\% of the barium isotopes and 100\% of scandium show HFS. 
This is in perfect agreement with the fact that the central peak of the barium 
signal is much higher than the other peaks, 
while the three peaks of the scandium signal have the same amplitude.}. 
However, as shown by \citet{Bel09c}, this is not the case: indeed the theory
unexpectedly fails to model this signal, even though the effects of HFS are 
correctly taken into account, and the same modeling assumptions used for barium 
are considered.
Such a failure has important consequences: if the Sc~{\sc ii} signal is 
produced by a physical mechanism neglected by \citet{Bel09c}, or that the 
theory does not account for (for example PRD effects), then not only it 
will be important to identify this mechanism, but it will be also 
interesting to understand why it seems not to be necessary in the case of 
barium.
Perhaps the problem on the physical origin of the triplet-peak structure 
of the Ba~{\sc ii} signal is still waiting for a complete answer!

\section{Theoretical approaches accounting for PRD effects}

The most natural way to describe PRD effects is by means of a 
{\it redistribution function}, a powerful tool which naturally appears in the 
``traditional'' scattering approach to the physics of resonance polarization
\citep[see, e.g.,][]{Ste94}.
This approach, which has been successfully applied for treating PRD phenomena, 
has, on the other hand, the limitation of being able to describe only 
two-level atoms with unpolarized, infinitely sharp lower level.
The redistribution function $\mathcal{R}$ is defined so that 
\begin{equation}
\mathcal{R}(\nu_2,\vec{\Omega}_2,\vec{e}_2;\nu_1,\vec{\Omega}_1,\vec{e}_1)
\, {\rm d}\nu_1 {\rm d}\nu_2 \frac{{\rm d} \Omega_1}{4 \pi} 
\frac{{\rm d} \Omega_2}{4 \pi}  \;  
\end{equation}
represents the probability that in a scattering process an incoming photon of 
frequency $\nu_1$, propagation direction $\vec{\Omega}_1$, and polarization 
state $\vec{e}_1$ is scattered in a photon with frequency $\nu_2$, propagation 
direction $\vec{\Omega}_2$, and polarization state $\vec{e}_2$. 
If the scattering of a radiation beam, whose polarization properties are 
described through the 4 Stokes parameters, is considered, then the 
redistribution function is replaced by a $4 \times 4$ redistribution matrix
${\mathcal R}_{ij}(\nu_2,\vec{\Omega}_2;\nu_1,\vec{\Omega}_1)$.

When the frequency dependence can be ignored, the redistribution matrix reduces 
to the so-called scattering phase matrix (or Rayleigh phase matrix)
${\mathcal P}_{ij}(\vec{\Omega}_2;\vec{\Omega}_1)$ 
\citep[see, e.g.,][or LL04]{Ham47,Ste94}.
In the atomic rest frame the angular and frequency dependencies can be 
separated, so that the redistribution matrix can be factorized into the 
product of the scattering phase matrix times a scalar function describing the 
frequency redistribution.
This is no more true in the laboratory frame where the Doppler effect, due to 
the motion of the atoms, has to be taken into account.
In this case, the scalar function describing the frequency redistribution also 
depends on the directions of the incoming and scattered beams, and an 
intricate coupling between frequency and angle dependencies appears 
\citep[see][for the unpolarized case]{Hum62}. 
In order to simplify the problem, the so-called hybrid approximation has often 
been used \citep[e.g.,][]{Ree82}. 
This approximation, which is justified only by heuristic arguments, consists 
in calculating the average over the angles of the scalar function describing 
the frequency redistribution, so that the angle and frequency dependencies can 
be factorized as in the atomic rest frame.

From the 1970s, the effects of PRD in frequency have been investigated using 
different approximate forms of the redistribution matrix, generally based on 
the angle averaged type I and type II redistribution functions of 
\citet{Hum62} \citep[see Sect.~1 of][for a historical review]{Sam10b}.
Particularly useful results were obtained by \citet{Fau87,Fau88} who
investigated the role of PRD effects considering both angle-averaged and 
angle-dependent redistribution functions, also including the effect of a 
magnetic field.
These investigations pointed out the limitations of approximate forms of the 
redistribution matrix previously proposed, and showed that the approximation 
of complete redistribution (CRD) is generally adequate to describe the line 
core polarization. 
The fact that PRD effects (i.e. coherent scattering) are important in the wings 
of strong resonance lines, while they are usually negligible in the line core 
has been then confirmed by several other investigations 
\citep[see, e.g.,][]{Fri96}.

From the theoretical point of view, the most tangled physical aspect in a 
scattering process of polarized radiation is the role of collisions (elastic 
and inelastic).
A first, fundamental investigation on this aspect was carried out by 
\citet{Omo72} who introduced the role of collisions in the quantum theory of 
Raman scattering developed by \citet{Fiu62}.
Starting from this work, \citet{Dom88} derived a very general analytic 
expression of the redistribution matrix for scattering of arbitrarily polarized 
radiation by an atom undergoing collisions. Using this expression \citet{Fau92} 
pointed out the sensitivity to the elastic collisional rate of the line-wing 
polarization, while \citet{Nag94} showed how depolarizing collisions affect 
the line-core polarization.
A detailed investigation on the role of various atmospheric parameters on 
the emergent linear polarization, calculated including PRD effects, 
has been recently carried out by \citet{Sam10b}.
To this end, \citet{Sam10a} developed new, very efficient numerical approaches 
for PRD radiative transfer based on the Gauss-Seidel and 
Successive-Overrelaxation iterative methods. 

Since the frequency redistribution matrix does not explicitly appear in the 
statistical equilibrium equations and in the radiative transfer equations,
in order to take into account PRD effects within the theoretical approach 
described in Sect.~3, it is necessary to consider the next terms in the 
perturbative expansion of the atom-radiation interaction (i.e. second and 
higher order processes). 
Unfortunately, such a generalization of the theory has revealed to involve many 
difficulties, since the equations that come out when higher order terms are 
included become immediately extremely complex and practically unmanageable,
unless further approximations are introduced.

Nevertheless, first important results have already been obtained.
Under the simplifying assumption of a two-level atom with unpolarized lower 
level, and neglecting stimulation effects, \citet{Bom97a} was able to perform 
all the calculations including higher order terms in Eq.~(\ref{eq:obs-evol}), 
thus deriving a generalized expression for the radiative transfer coefficients.
With respect to the expression obtained when only first-order processes are 
taken into account (valid under the complete redistribution approximation), 
the new emission coefficient contains an extra term (the so-called Rayleigh 
scattering term) which describes the contribution of coherent scattering.
Remarkably, the redistribution matrix of \citet{Dom88} is recovered if the 
generalized emission coefficient derived by \citet{Bom97a} is expressed in 
terms of a redistribution matrix:
\begin{eqnarray}
	\varepsilon_i(\nu_2,\vec{\Omega}_2) \!\!\!\! & = 
	& \!\!\! \eta^{(0)} 
	\frac{\nu_2^4}{\nu_0^4} \int {\rm d}\nu_1 
	\oint \frac{{\rm d}\Omega_1}{4 \pi} \sum_{j=0}^3
	R_{ij}(\nu_2, \vec{\Omega}_2; \nu_1, \vec{\Omega}_1) \, 
	S_{\!\! j \,}(\nu_1,\vec{\Omega}_1) \nonumber \\
	& & \!\!\! + \eta^{(0)} \frac{v_2^4}{\nu_0^4} \,
	\epsilon \, B_P(\nu_0) \, \phi(\nu_0-\nu) \, \delta_{i,0}  \; ,
\end{eqnarray}
with $i=0,1,2,3$ (corresponding respectively to the Stokes parameters {\it I,
Q, U}, and {\it V}), $\eta^{(0)}$ the line integrated absorption coefficient, 
$\nu_0$ the line frequency, and $S_{\!\! j \,}$ ($j=0,\cdots,3$) the four 
Stokes parameters.
The last term describes the contribution to the emitted radiation coming from 
collisionally excited atoms, the branching ratio being 
$\epsilon=\frac{\Gamma_I}{\Gamma_I+\Gamma_R}$, with $\Gamma_I$ and $\Gamma_R$ 
the inelastic collisions de-excitation rate and the radiative de-excitation 
rate, respectively.
The $\delta_{i,0}$ shows that this contribution is unpolarized (collisions are 
assumed to be isotropic). The quantity $B_P$ represents the Planck function in 
the Wien limit, while $\phi(\nu_0-\nu)$ is a normalized Lorentzian profile.
The redistribution matrix is given by \citep[see][]{Bom97b}
\begin{equation}
\label{eq:red-mat}
	R_{ij}= \sum_{K=0}^2 W_K(J_{\ell}, J_u) \, \Big[ {\mathcal P}^{(K)}
	(\vec{\Omega}_2; \vec{\Omega}_1) \Big]_{ij} \, \Big\{ \alpha \,
	r_{II}(\nu_2,\nu_1) + \Big[ \beta^{(K)} - \alpha \Big] \,
	r_{III}(\nu_2, \nu_1) \Big\}  \; .
\end{equation}
This expression is valid in the atomic rest frame, where, as can be 
observed, the frequency and angular dependencies can be factorized. 
The quantity $W_K$ is a numerical factor introduced by \citet{Lan84}, 
the matrix ${\mathcal P}^{(K)}$ is the $K$-th multipole component of the 
Rayleigh phase scattering matrix, also introduced by \citet{Lan84}. 
The functions $r_{II}(\nu_2,\nu_1)$ and $r_{III}(\nu_2,\nu_1)$ are given by 
\begin{eqnarray}
	r_{II}(\nu_2,\nu_1) & \!\! = \!\! & \phi(\nu_0-\nu_1) \, 
	\delta(\nu_2-\nu_1)  \; , \\
	r_{III}(\nu_2,\nu_1) & \!\! = \!\! & \phi(\nu_0-\nu_1) \, 
	\phi(\nu_2-\nu_1)  \; ,
\end{eqnarray}
where $\phi(\nu_0-\nu_1)$ is a Lorentzian absorption profile, while 
$\delta(\nu_2-\nu_1)$ is the Dirac delta.
The first function describes the coherent contribution to the scattering, the 
second one the contribution in complete redistribution.
The branching ratios are given by $\alpha$ and 
$\big[ \beta^{(K)} - \alpha \big]$, respectively, with
\begin{equation}
	\alpha = \frac{\Gamma_R}{\Gamma_R + \Gamma_I + \Gamma_E} 
	 \; , \;\;\;\; {\rm and} \;\;\;\;
	\beta^{(K)} = \frac{\Gamma_R}{\Gamma_R + \Gamma_I + D^{(K)}}  \; , 
\end{equation}
where $\Gamma_E$ is the elastic collisions rate (responsible for 
line-broadening and for destruction of frequency correlations between incoming 
and scattered photons), and $D^{(K)}$ is the $K$-multipole depolarizing rate 
due to elastic collisions (responsible for depolarization).\\
The quantity $\alpha$ is the probability that a radiative decay takes place 
before any kind of collision: it thus represents the branching ratio for the 
scattering processes that are coherent in the atom rest frame.
The quantity $\beta^{(K)}$, on the other hand, is the probability that a 
radiative decay occurs before an elastic depolarizing collision.
The difference $\big[ \beta^{(K)} - \alpha \big]$ thus gives the probability 
that the radiative decay takes place after the atom undergoes an elastic 
collision that redistributes the photon frequency, but that does not destroy 
atomic polarization: it thus represents the branching ratio for the scattering
processes that are completely redistributed in frequency in the atom rest 
frame.
It should be observed that a branching ratio term ($1- \epsilon$), taking into 
account the fraction of atoms that are radiatively excited, is already 
included in the branching ratios $\alpha$ and $\big[ \beta^{(K)}-\alpha \big]$
\citep[for a discussion on the physical meaning of the various 
quantities appearing in Eq.~(\ref{eq:red-mat}), see also][]{Sam10b}.
In this context, the CRD limit is obtained when the elastic collisions are so 
efficient that any frequency correlation between the incoming and outgoing 
photons is completely destroyed.
Indeed, when $\Gamma_E \! >> \! \Gamma_R$ the coherent term disappears 
($\alpha \rightarrow 0$), and the resulting redistribution matrix corresponds 
to the one which can be obtained within the theoretical framework described in 
Sect.~3 (valid under the flat spectrum approximation).
As already observed in Sect.~3, when the elastic collisional rate is very high,
depolarizing collisions are also very efficient, and polarization phenomena 
become negligible. 
For this reason, when the CRD limit due to high collisional rates is 
considered, it is customary to assume the rates $\Gamma_E$ and $D^{(K)}$ as 
independent, though they have the same physical origin.
On the other hand, in the limit of no collisions ($\Gamma_E = D^{(K)} = 0$)
the branching ratio $\big[ \beta^{(K)} - \alpha \big]$ vanishes, and the 
scattering is completely coherent, as implied by energy conservation.

Within the same approach, \citet{Bom97b} also investigated the 
effect of a magnetic field of arbitrary strength (which is found to mix 
angle and frequency dependencies also in the atom rest frame), as well as the 
transition to the laboratory frame.

Another interesting theoretical approach, based on the density matrix 
formalism, which can be applied to multi-level atoms, in the presence of 
arbitrary magnetic fields, and that does not require the flat spectrum 
approximation to be assumed, has been proposed by \citet{Lan97}. 
In this approach the density-matrix formalism is generalized to the framework 
of the so-called metalevels theory, in which the atomic levels are assumed to 
be composed by a continuous distribution of infinitely sharp sublevels.
The formulation presented in \citet{Lan97} does not take collisions into 
account, and thus describes the limit of pure coherent scattering.
This scheme has been applied by \citet{Lan98} for the interpretation of the 
peculiar signal observed in the second solar spectrum of the sodium doublet.

More recently, ``intermediate'' theoretical schemes, exploiting the strength 
points both of the scattering approach, and of the density matrix formalism 
have been proposed. 
An interesting example is the approach proposed by \citet{Hol05} for the 
investigation of the physical origin of the triplet-peak structure shown by 
several signals of the second solar spectrum.
Although these approaches are able to account for many fundamental physical 
aspects, in general they are not derived in a self-consistent way from basic 
physical principles. 
This makes extremely important, and in some cases also very difficult, to 
clearly establish their limits of applicability, which might often result to 
be fairly restrictive.

\section{Conclusions}
While until the end of the 1990s, the various branches of solar polarimetry, 
from the instrumentation to the observational techniques, from the theory of 
polarization to its solar application, were developed hand in hand, in the 
last ten years, as it can also be noticed from the references mentioned in 
this paper, progresses in the field of pure theory are proceeding significantly 
slower than in the other research branches.

In 1995, when the first ``Solar Polarization Workshop'' was organized, 
the need of developing theoretical tools suitable for describing the effects 
of frequency redistribution was already clear, and the problem of deriving a 
general analytical form of the redistribution function was strongly 
debated. 
Today, fifteen years later, our understanding of this physical aspect has 
substantially improved, different promising approaches have been proposed, 
and for the case of a two-level atom (with unpolarized lower level) the PRD 
problem seems to be solved.
Nevertheless, the development of a self-consistent, general theory for 
radiative transfer in the presence of PRD effects (applicable to multi-levels 
atoms, and accounting for the fundamental role of quantum coherences) remains 
since many years a fundamental unsolved problem.
We are aware of the difficulties that the formulation of such a theory 
implies, and it is reasonable to expect that its development will proceed 
rather slowly, through a series of small, demanding steps.
The strong possibility of not achieving novel and outstanding results in short 
or medium periods might explain why not many people (in particular among the 
new generations) decide to cope with this difficult challenge, definitely 
necessary for a new substantial acceleration in this research field.

The theory developed by Landi Degl'Innocenti at the beginning of the 1980s, 
and explained in great detail in LL04, keeps representing the most solid, 
advanced and complete theoretical approach to the physics of spectral line 
polarization developed so far.
More than twenty years after its first development, after a countless 
number of successful applications, its soundness for the interpretation of all 
those signals which can be treated within the limit of complete frequency 
redistribution appears to be well established.

\acknowledgements Finantial support by the Spanish Ministry of Science through
project AYA2010-18029 (Solar Magnetism and Astrophysical Spectropolarimetry)
is gratefully acknowledged.

\end{document}